\documentclass[aps,prb,preprint]{revtex4-1}
\usepackage{graphicx}
\usepackage{epstopdf}
\usepackage{tabularx}

\usepackage{bm,amsmath,amssymb}
\usepackage{color, soul}
\usepackage{xcolor}
\usepackage{dcolumn} 
\usepackage{multirow}

\begin{document}

\title{Deep Learning of Accurate Force Field of Ferroelectric HfO$_2$ }

\author{Jing Wu}
\affiliation{Fudan University, Shanghai 200433, China}
\affiliation{School of Science, Westlake University, Hangzhou, Zhejiang 310024, China}
\affiliation{Institute of Natural Sciences, Westlake Institute for Advanced Study,Hangzhou, Zhejiang 310024, China}
\author{Yuzhi Zhang}
\affiliation{ Yuanpei College, Peking University, Beijing 100871, China}
\author{Linfeng Zhang}
\affiliation{Program in Applied and Computational Mathematics, Princeton University, Princeton, NJ 08544, USA}
\author{Shi Liu}
\email{liushi@westlake.edu.cn}
\affiliation{School of Science, Westlake University, Hangzhou, Zhejiang 310024, China}
\affiliation{Institute of Natural Sciences, Westlake Institute for Advanced Study,Hangzhou, Zhejiang 310024, China}

\date{\today}

\begin{abstract}{
The discovery of ferroelectricity in HfO$_2$-based thin films opens up new opportunities for using this silicon-compatible ferroelectric to realize low-power logic circuits and high-density non-volatile memories. The functional performances of ferroelectrics are intimately related to their dynamic responses to external stimuli such as electric fields at finite temperatures. Molecular dynamics is an ideal technique for investigating dynamical processes on large length and time scales, though its applications to new materials is often hindered by the limited availability and accuracy of classical force fields. Here we present a deep neural network-based interatomic force field of HfO$_2$ learned from {\em ab initio} data using a concurrent learning procedure. The model potential is able to predict structural properties such as elastic constants, equation of states, phonon dispersion relationships, and phase transition barriers of various hafnia polymorphs with accuracy comparable with density functional theory calculations. The validity of this model potential is further confirmed by the reproduction of experimental sequences of temperature-driven ferroelectric-paraelectric phase transitions of HfO$_2$ with isobaric-isothermal ensemble molecular dynamics simulations. We suggest a general approach to extend the model potential of HfO$_2$ to related material systems including dopants and defects. 
}
\end{abstract}

\maketitle

\newpage
\section{Introductions}
Ferroelectrics characterized by the electric field-tunable polarization, fast switching speed, low power consumption, and high endurance have been considered as excellent materials to realize high-speed energy-efficient logic and nonvolatile memory devices~\cite{Scott07p954,Huang18p1700560,Mikolajick20p1434}. However, the poor compatibility of conventional perovskite ferroelectrics such as Pb(Zr, Ti)O$_3$ with the complementary metal-oxide-semiconductor (CMOS) technology has made it difficult to down scale the ferroelectric memory to the sub-100 nm regime~\cite{Pinnow04pK13}. Though the first commercial ferroelectric random-access memory (FeRAM) appeared in the early 1990s~\cite{Bondurant90p273}, current state-of-art technology node remains 130 nm~\cite{McAdams04p667}. In comparison, silicon-based memories such as DRAM and NAND flash memory have already achieved the 10-nm technology node, delivering much lower cost per bit than FeRAM~\cite{Park18p795}. Finding ferroelectrics with improved CMOS compatibility thus becomes a key task for the development of ferroelectric memory technology~\cite{Pesic17p1236,Park18p795}.  

The discovery of robust nanoscale ferroelectricity in HfO$_2$-based thin films by \texttt{NamLab} in 2011 opened up exciting opportunities for ferroelectric-based electronics~\cite{Boscke11p102903}. Hafnium oxide, being thermodynamically stable on silicon, has proved CMOS compatibility~\cite{Gutowski02pB3.2} and is already used as the high-permittivity gate insulator in silicon-based field effect transistors. Experimentally, it was found that an ultrathin doped HfO$_2$ film of just $\approx$1~nm can still support switchable polarization~\cite{Cheema20p478}, free from the depolarization effect often presented in thin films of perovskite ferroelectrics~\cite{Batra73p3257,Wurfel73p5126,Ma02p386}. Moreover, current atomic layer deposition (ALD) technique is capable of depositing hafnium oxides in high-aspect-ratio structures on silicon, allowing the fabrication of high-quality 3D-stackable memory~\cite{Park18p795,Luo20p1391}. 

The origin of ferroelectricity in HfO$_2$-based thin films has been an active research topic since its discovery. Despite the relatively simple chemical composition, HfO$_2$ is known to form many polymorphs. At room temperature, bulk HfO$_2$ will crystallize in a monoclinic phase ($m$-phase, space group $P2_1/c$), which evolves to a tetragonal phase ($t$-phase) of space group $P4_2/nmc$ and subsequently to a cubic phase of space group $Fm\bar{3}m$ with increasing temperature.  The phase transition of HfO$_2$ at ambient temperature with increasing pressure follows $P2_1/c \rightarrow Pbca \rightarrow Pnma$~\cite{Ohtaka04p1369}. All these polymorphs have inversion symmetry thus forbidding spontaneous polarization. Combined experimental and theoretical studies eventually pinpointed the phase responsible for the ferroelectricity: an orthorhombic phase in the space group of $Pca2_1$ ($po$-phase)~\cite{Park15p1811, Huan14p064111,Sebastian14p140103,Sang15p162905,Materlik15p134109}. However, a series of first-principles density functional theory (DFT) studies revealed that the $po$-phase has energy higher than the $m$-phase~\cite{Huan14p064111,Sebastian14p140103,Materlik15p134109}, whereas simply applying hydrostatic pressures or epitaxial strains is not enough to make the $po$-phase favored over the $m$-phase~\cite{Sebastian14p140103,Batra17p4139}. The general consensus now is that the thermodynamic stability of ferroelectric HfO$_2$ results from combined effects of various factors such as doping~\cite{Schroeder14p08LE02, Starschich17p333,Park17p4677,Xu17p124104,Batra17p9102}, mechanical stress~\cite{Shiraishi16p262904,Batra17p4139}, oxygen vacancy~\cite{Xu16p091501,Pal17p022903}, surface/interface/grain boundary energy~\cite{Materlik15p134109, Park15p1811, Polakowski15p232905, Batra16p172902,Knneth17p205304,Park17p9973}, electric fields~\cite{Batra17p4139}, and substrate orientations~\cite{Liu19p054404}. More recently, it was pointed out that the flat polar phonon bands in HfO$_2$ give rise to intrinsically localized dipoles, responsible for the robust scale-free ferroelectricity~\cite{Lee20p1343}. 

Like all other ferroelectrics, the functionalities of HfO$_2$-based ferroelectrics depend on various kinetic and dynamical processes that often span many scales in time and space. Recent experiments suggest that the thermodynamic arguments are not enough to explain the emergence of the metastable $po$-phase~\cite{Park17p9973, Park18p716}. Park {\em et al.} found that the fraction of the low-entropy $m$-phase in Hf$_{0.5}$Zr$_{0.5}$O$_2$ thin films increases with increasing temperature, contradictory to the prediction of the thermodynamic model that a higher temperature will favor high-entropy phases such as $t$-phase and $po$-phase~\cite{Park17p9973}. The kinetic effect of phase transitions during the annealing and cooling processes likely contribute to the formation of the polar phase~\cite{Park18p716,Park18p1800522}. Polarization switching is another important dynamical process for a ferroelectric as the switching speed and coercive field dictate the writing speed and power consumption, respectively~\cite{Liu16p360,Li19p126502}. However, the atomistic mechanisms and characteristics of ferroelectric switching in this fluorite-structure ferroelectric remain largely unexplored, while experimental measurements reported in literature seem to support different switching mechanisms~\cite{Gong18p262903,Yoon19p050601,Hoffmann19p464}. Therefore, it is desirable to have a tool to study the kinetic and dynamical properties of HfO$_2$-based ferroelectrics at the atomic level. 

First-principles DFT calculations have played an important role in understanding the structure-property relationship of ferroelectrics. Nevertheless, the study of finite-temperature dynamical properties of ferroelectrics is still beyond the reaches of conventional DFT methods due to the expensive computational cost. Statistical methods such as molecular dynamics (MD) simulations are ideal techniques for investigating dynamical processes on larger length/time scales while providing atomistic details with femtosecond time-resolution. In the case of HfO$_2$, several force fields have already been developed~\cite{Shan10p125328,Broglia14p065006,Schie17p094508,Sivaraman20p104}. However, none of them considered the ferroelectric $Pca2_1$ phase during the parameterization, and it is not yet clear whether those force fields can accurately describe the structural properties of the ferroelectric phase. Such situation also reveals the limitation of MD simulations: applications to new materials systems are often hindered by the limited availability and accuracy of classical force fields. Developing a force field is often a tedious process because of the many-body nature of the potential energy. Most force fields approximate the interatomic interactions with sets of relatively simple analytical functions in which the parameters are fitted to a database of information including quantum mechanical calculations and/or experimental thermodynamic properties. The ``true" interatomic potential of complex materials is intrinsically a high-dimensional function, which can only be roughly approximated by analytical functions of ``ad hoc" forms with a limited number of parameters. Moreover, the transition metal-oxygen bonds in ferroelectrics often possess a mixed ionic-covalent character~\cite{Cohen92p136} due to the $p$-$d$ hybridization, making the force field development even more challenging~\cite{Phillpot07p239, Liu13p104102,Liu13p102202}.

The application of machine learning (ML) to force field development offers an attractive solution to the accuracy-efficiency dilemma by combining the strengths of DFT and classical MD. Many ML-based force fields have been developed for systems of vastly different bonding characters, ranging from organic molecules~\cite{Smith17p3192,Manzhos08p224104}, molecular and condensed water~\cite{Bartok13p054104,Morawietz16p8368,Ko19p3269}, to metals~\cite{Eshet10p184107,Botu15p094306,Zhang20p107206} and alloys~\cite{Andolina20p154701,Zhang19p023804,Jiang2020arXiv}, semiconductors such as silicon~\cite{Sanville08p285219,Behler08p2618,Babaei19p074603,Bartok18p041048} and GeTe~\cite{Sosso12p174103}, and to inorganic halide perovskites~\cite{Thomas19p134101}. 
In general, there are two key ingredients in a ML-based force field: a {\em descriptor} that represents the local atomic environment and a non-linear 
{\em fitting function} that maps the descriptor to the local energy contribution.
For example, Behler and Parrinello (BP) proposed to use ``symmetry functions" to describe the local geometric environment of an atom, which were then used as inputs for an artificial neural network (NN) to evaluate the atomic contribution to the total energy~\cite{Behler07p146401}. 
Bartok~{\em et al.} developed a Gaussian approximation potential (GAP) for silicon using the smooth overlap of atomic positions (SOAP) kernel~\cite{Bartok13p184115} that quantifies the similarity between atomic neighborhoods characterized by neighbor densities~\cite{Bartok18p041048}. 
More recently, the smooth edition of the Deep Potential (DP) scheme~\cite{Zhang18p143001,Zhang18p4436} employed a faithful and symmetry-preserving embedding network to parametrize the descriptors, bypassing the need to fix hand-crafted descriptors and enabling an end-to-end procedure for representing complex chemical environments in chemical reactions~\cite{Zeng2019arXiv}, heterogeneous aqueous interfaces~\cite{Andrade20p2335}, and high-entropy alloys~\cite{Dai20p168}. 

In this work, we applied the Deep Potential Molecular Dynamics (DeePMD) method~\cite{Zhang18p143001,Zhang18p4436} to construct an accurate and transferable force field for HfO$_2$ by concurrently learning from results of DFT calculations~\cite{Zhang19p023804,Zhang20p107206}. The resultant DP model reproduces the DFT results of a wide range of thermodynamic properties of various hafnia polymorphs, including the ferroelectric $Pca2_1$ phase. 
Notably, the temperature-driven ferroelectric-paraelectric phase transition of HfO$_2$ is well captured by MD simulations in the isobaric-isothermal ($NPT$) ensemble. The DP predictions of transition barriers between different phases of HfO$_2$ ($P2_1/c$, $Pca2_1$, $Pbca$, and $P4_2/nmc$) agree well with first-principles results. We believe current DP model of HfO$_2$ can be systematically improved and extended by adding new training data representing new atomic environments, enabling atomistic modeling of various extrinsic effects such as doping and defects. 

 \section{Computational Methods}
\subsection{ Deep potential molecular dynamics }
We briefly discuss the key concepts in DeePMD method and refer interested readers to the original papers~\cite{Zhang18p143001,Zhang18p4436} for detailed discussions. The DP model assumes the total potential energy ($E$) can be expressed as a sum of atomic energies ($E^i$), $E = \sum_iE^i$. Each atomic energy $E^i$ is parameterized with a deep neural network (DNN) function defined as $E^i = E^{{\mathbf \omega}_{\alpha_i}}(\bm {\mathcal{R}}^i)$, where $\bm {\mathcal{R}}^i$ is the local environment of atom $i$ in terms of Cartesian coordinates relative to its neighbors within a cutoff radius $r_c$, $\alpha_i$ denotes the chemical species of $i$th atom, and ${\mathbf \omega}_{\alpha_i}$ is the DNN parameter set that eventually will be optimized by the training procedure. It is noted that each sub-network of $E_i$ consists of an embedding and a fitting neural network. 
The embedding network maps $\bm {\mathcal{R}}^i$ to a feature matrix  $\bm {\mathcal{D}}^i$ that preserves the permutation, translation, and rotation symmetries of the system, while the fitting network is a standard feedforward neural network that maps $\bm {\mathcal{D}}^i$ to the atomic energy $E^i$. 

In this work, the smooth version of the DP model was employed~\cite{Zhang18p4436} and the \texttt{DeePMD-kit} package~\cite{Wang18p178} was used for training. 
The cut-off radius is set to 6 \AA, and the inverse distance $1/r$ decays smoothly from 1 \AA ~to 6 \AA~to remove the discontinuity introduced by the cut-off. The embedding 
network of size (25, 50, 100) follows the ResNet-like architecture. The fitting network is composed of three layers, each containing 240 nodes.
As reported in ref~\cite{Zhang18p143001}, the loss function is defined as 
\begin{equation}
L({p}_\epsilon, {p}_f, {p}_\xi) = {p}_\epsilon \Delta{\epsilon}^2 + \frac{p_f}{3N} \sum_i \left| \Delta{{\bm {F}}}_i \right|  + \frac{p_\xi}{9}  \left \| \Delta \xi \right \|^2
\end{equation}
where $\Delta$ denotes the difference between the DP prediction and the training data, $N$ is the number of atoms, $\epsilon$ is the energy per atom, ${\bm {F}}_i$ is the atomic force of atom $i$, and $\xi$ is the virial tensor divided by $N$. $p_\epsilon$, $p_f$, and $p_\xi$ are tunable prefactors. Here we increase both $p_\epsilon$ and $p_\xi$ from 0.02 to 1. And  $p_f$ decrease from 1000 to 1. 

We note here that the additive structure $E = \sum_iE^i$ is an {\em ansatz} of the DP model, and of many other ML-based force fields. Such an ansatz ensures that the potential energy is extensive, so that the same model can be used to describe systems with different number of atoms. The introduction of the cutoff radius $r_c$ makes the interaction range finite and potentially misses some long-range effect. On the other hand, in many cases, the finite-range model indeed gives an accuracy of $\sim1$meV/atom in energy, which is comparable with the intrinsic error of the functional approximation adopted in DFT, and is sufficient for most properties of practical interest. This is indeed the case for the system we study here. The incorporation of dopants and defects, as well as finite fields, may require a more delicate treatment of the long-range interactions, which will be left to future investigations.

\subsection{Deep potential generator }
\begin{figure}[ht]
\centering
\includegraphics[scale=0.2]{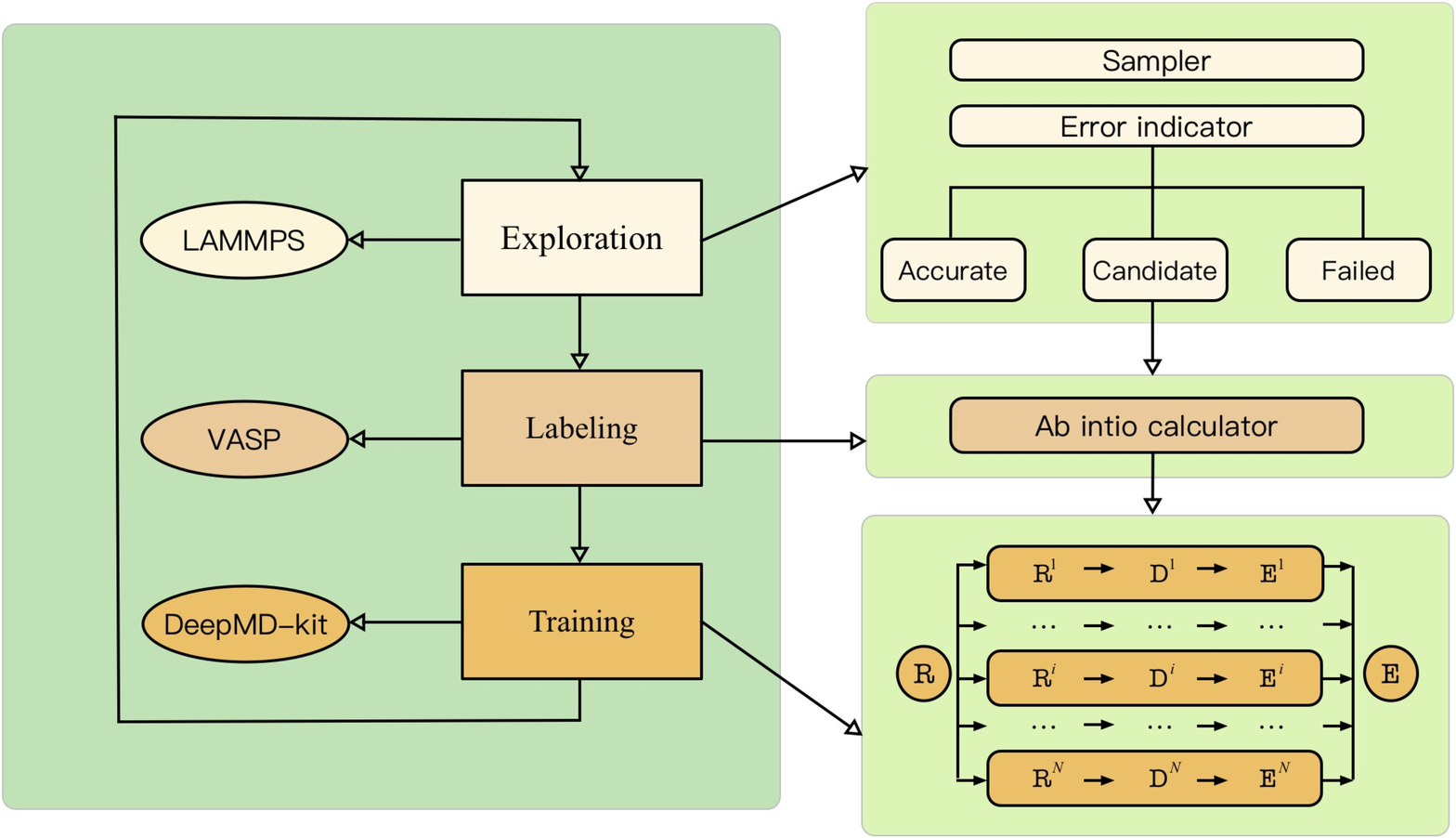}
 \caption{DP-GEN workflow. One cycle of DP-GEN contains three steps: exploration, labeling, and training. Molecular dynamics simulations using a DP model are performed in the exploration step to sample new configurations, among which candidate configurations are selected by the error indicator. The labeling step undertakes {\em ab initio} calculations for the candidate configurations obtained in the exploration step. An ensemble of new DP models are then re-trained using the updated training database.}
  \label{workflow}
 \end{figure}

Since {\it ab initio} calculations are expensive, 
to develop a reliable ML-based potential, 
we need a procedure that generates an optimal and minimal set of training data that covers a wide range of relevant configurational space.
Here we employ the Deep Potential generator (DP-GEN) scheme~\cite{Zhang19p023804}. 
DP-GEN is a concurrent learning procedure involving three steps, {\em exploration, labeling, and training,}  which form a closed loop (Fig.~\ref{workflow}). 
Starting with an {\em ab initio} database, an ensemble of DP models are trained with different initial values of $\mathbf{\omega}_{\alpha_i}$. 
In the {\em exploration} step, one of these models is used for MD simulations to explore the configuration space. For each newly sampled configuration from MD, the ensemble of DP models will generate an ensemble of predictions ({\em e.g.}, energies and atomic forces). Since the ensemble of models only differ in the initialization of network parameters $\mathbf{\omega}_{\alpha_i}$, these models will exhibit nearly identical predictive accuracy for configurations that are well represented by the training data. Otherwise, they are expected to give scattered predictions with a considerable variance. 
Therefore, the deviation of the model predictions can be used to formulate the criterion for {\em labeling}: a sampled configuration giving rise to a large model deviation will be labeled via DFT calculations and will be added to the training database for {\em training} in the next cycle. 

In detail, the model deviation $\mathcal{E}$ is defined as the maximum standard deviation of the predictions of the atomic forces $\bm{F}_i$,
\begin{equation}
\mathcal{E} = \max_i \sqrt {\left < || \bm{F}_i-  \left <  \bm{F}_i    \right > ||^2 \right > }
\end{equation}
where $\left< ...\right>$ is the average taken over the ensemble of DP models. 
In practice, we introduce two thresholds, $\mathcal{E}_{\rm lo}$ and $\mathcal{E}_{\rm hi}$. 
Only configurations satisfying $\mathcal{E}_{\rm lo} < \mathcal{E} < \mathcal{E}_{\rm hi}$ are labeled for DFT calculations, because a configuration with a small $\mathcal{E} < \mathcal{E}_{\rm lo}$ is already well described by the current DP model, whereas a configuration with a large model deviation $\mathcal{E} > \mathcal{E}_{\rm hi}$ is likely to be highly distorted or unconverged in DFT calculations. 

When all sampled configurations have $\mathcal{E}< \mathcal{E}_{\rm lo}$, the ensemble of DP models is considered converged. Here we set $\mathcal{E}_{\rm lo}=0.12$~eV/\AA~and $\mathcal{E}_{\rm hi}=0.25$~eV/\AA. The described automatic and iterative workflow was performed using \texttt{DP-GEN} package.

\subsection{Initial training database and exploration protocol}
The initial training database contains structures generated by randomly perturbing ground-state structures of  $P2_1/c$, $Pbca$,  $Pca2_1$ and $P4_2/nmc$ phases of HfO$_2$. 
We use $2 \times2 \times 2$ supercells of 96 atoms for DFT calculations with the Vienna Ab initio Simulation (\texttt{VASP}) package~\cite{Kresse96p11169,Kresse96p15}. 
The projected augmented wave (PAW) method~\cite{Blochi94p17953,Kresse99p1758} and the generalized gradient approximation of Perdew-Burke-Ernzerhof (PBE)~\cite{Perdew96p3865} type for the exchange-correlation functional are employed. 
An energy cutoff of 600 eV and $2\times2\times2$ $k$-grid mesh are sufficient to converge the energy and atomic force.  At the exploration step, the configuration space is sampled by running $NPT$ simulations at various temperatures (from 100 to 3300~K) and pressures (from $-50$ to $400$~kBar). Because the training database will keep incorporating new configurations generated and labeled on the fly during the exploration, we expect the final converged DP model is not sensitive to the exact construction of the initial training database.

\subsection{MD simulations of phase transition} 
The optimized DP model of HfO$_2$ is used to study phase transitions driven by temperature by performing $NPT$ MD simulations. We use a $8 \times8 \times 8$ supercell of 6144 atoms and a time step of 1~fs. The temperature is controlled via the Nos\'e-Hoover thermostat and the pressure is maintained using the Parrinello-Rahman barostat as implemented in \texttt{LAMMPS}~\cite{Plimpton95p1}. The final configuration of the simulation at a lower temperature is used as the initial configuration for the simulation at a higher temperature. 

 \section{Results and Discussions}
\subsection{Fitting performance of DP model}
\begin{figure}[ht]
\centering
\includegraphics[scale=1.0]{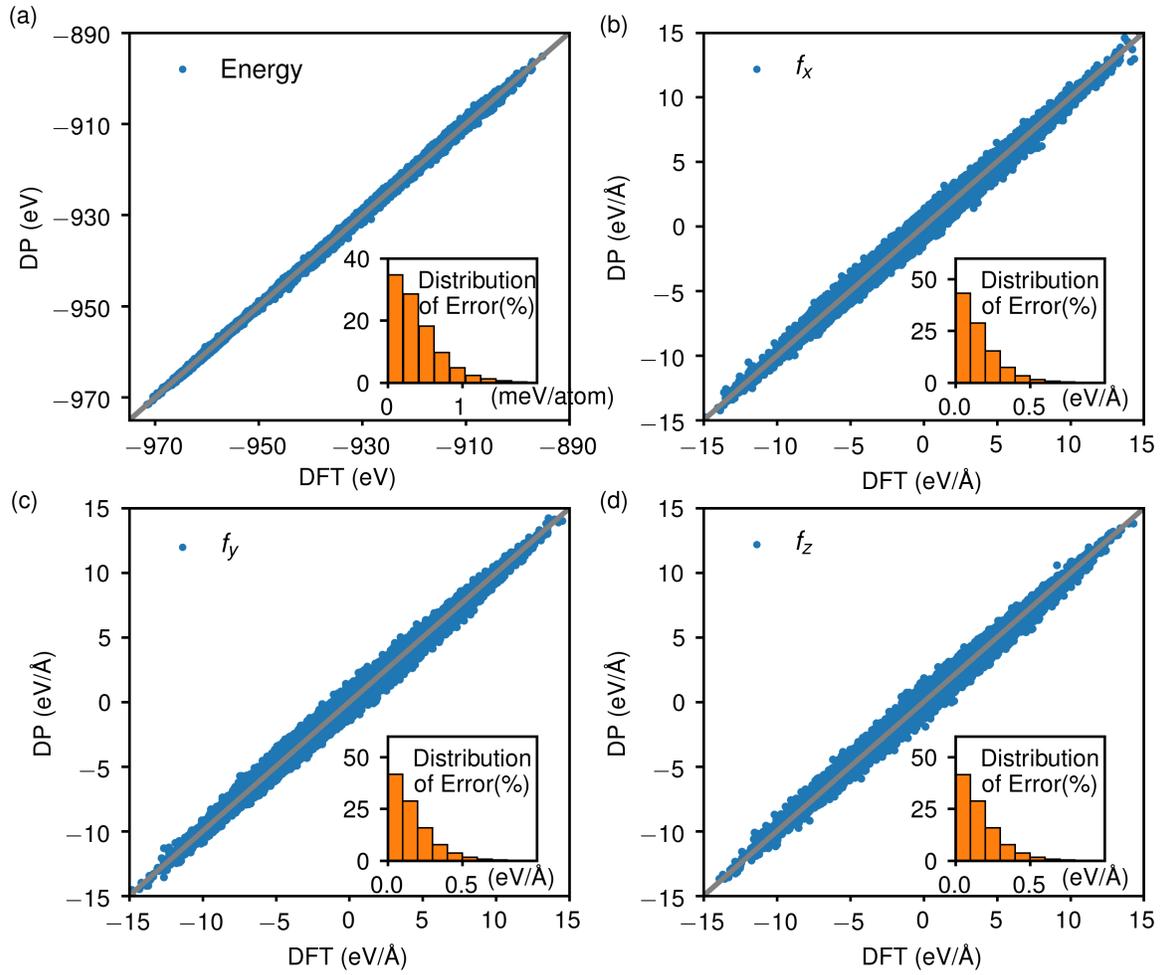}
 \caption{ Comparison of (a) energies and (b-d) atomic forces predicted using the DP model with reference DFT results for configurations in the final training database. The insets provide the distribution of the absolute error.}
  \label{fit}
 \end{figure}

Figure~\ref{fit} compares the energies and atomic forces predicted by DFT and DP for all the structures in the final training database (21768 configurations) with insets showing the distributions of absolute errors. We find an overall satisfactory agreement between DP predictions and DFT results with a mean absolute error (MAE) of 1.6 meV/atom for energy. This clearly demonstrates that the deep neural network-based potential model has excellent representability, capable of learning complex and highly non-linear energy functional with little human intervention. The whole DP-GEN process carried out 61 iterations during which a total number of 41 million configurations were sampled with only 21768 (0.05$\%$) configurations selected for labeling. The usage of model deviation as an error indicator for labeling substantially reduced the computational cost associated with DFT calculations.  

\subsection{Predictions of static properties of hafnia polymorphs }
Table~\ref{latt-param} compares the lattice parameters of different phases of HfO$_2$ optimized with DFT and DP at 0~K, demonstrating excellent agreement. Elastic constants and moduli are fundamental material properties as they reflect the strength of chemical bonds that are intimately related to the second derivative of the potential energy.  We use the DP model to calculate the elastic properties for a few hafnia polymorphs,  $P2_1/c$, $Pbca$, $Pca2_1$, $P4_2/nmc$, $Fm\bar{3}m$, $P2_12_12$, $Pbcn$, and $Pmn2_1$, and compare the DP values with DFT results. 

\begin{figure}[ht]
\centering
\includegraphics[scale=1.5]{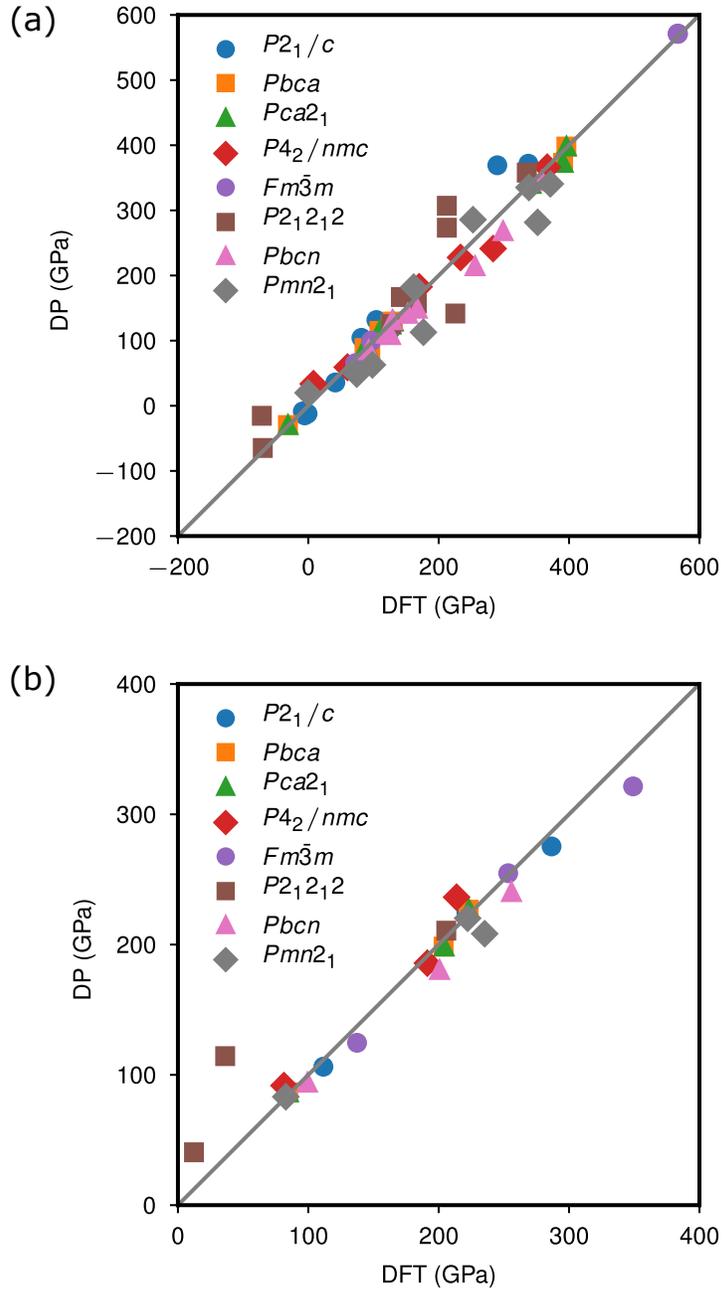}
 \caption{(a) Elastic constants and (b) various moduli (shear, bulk, and Young's modulus) of different HfO$_2$ polymorphs. }
  \label{elastic}
 \end{figure}

\begin{figure}[ht]
\centering
\includegraphics[scale=1.5]{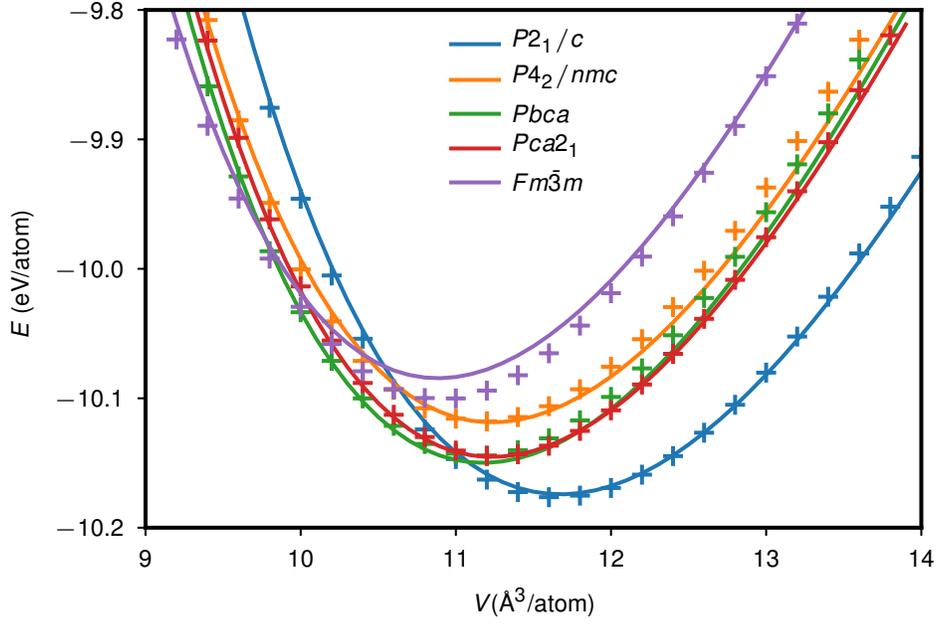}
 \caption{Equation of states of different HfO$_2$ polymorphs. Solid lines and cross points denote DFT and DP results, respectively. The DP model predicts the correct sequence of phase stability: $E(P2_1/c) < E(Pbca)<  E(Pca2_1) < E(P4_2/nmc) < E(Fm\bar{3}m)$.}
  \label{eos}
 \end{figure}

As illustrated in Fig.~\ref{elastic} and detailed in Table~\ref{Telas}, the elastic constants and moduli from the DP model are comparable with the DFT values. Considering that the values of elastic constants distribute over a wide range from $-50$~GPa to 600 GPa, the demonstrated agreement between DP and DFT results highlight the accuracy of the optimized model. It is noted that the training database does not contain any elastic property nor any structural information of $P2_12_12$, $Pbcn$, and $Pmn21$ phases explicitly. The ability of the DP model to predict reasonably well the elastic properties of phases not included in the training database with quantum mechanical accuracy highlights its accuracy as well as transferability. DP and DFT predictions of equations of states (EoSs) of selective hafnia polymorphs are reported in Fig.~\ref{eos}. It is clear that DP well reproduce DFT EoSs as well as the order of phase stability: $E(P2_1/c) < E(Pbca)<  E(Pca2_1) < E(P4_2/nmc) < E(Fm\bar{3}m)$. It is remarkable that DP is capable of capturing the small energy difference between $Pbca$ and $Pca2_1$. 

\begin{figure}[ht]
\centering
\includegraphics[scale=0.25]{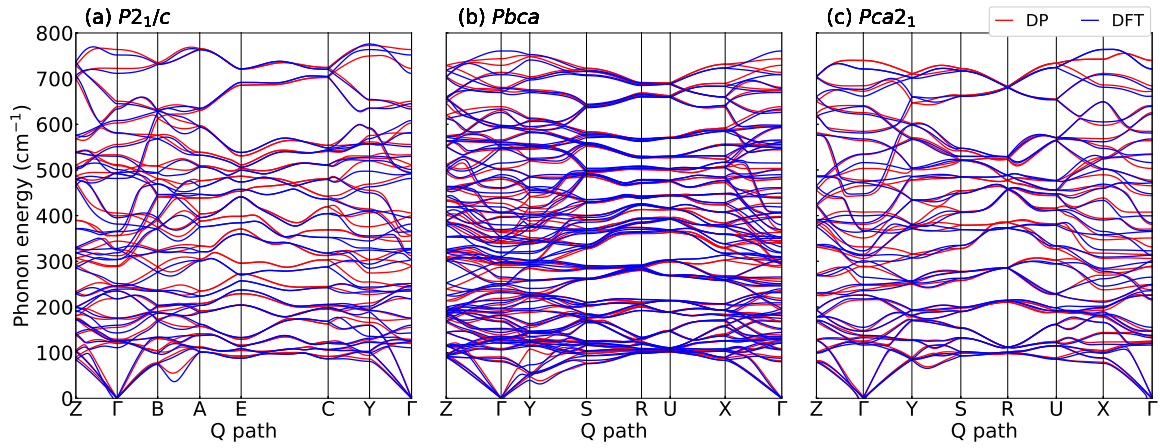}
 \caption{\label{phonon}
 The phonon dispersion relations of three different phases of HfO$_2$. The \texttt{phonopy} package~\cite{Togo15p1} was used to produce the results.}
 \end{figure}

To further investigate the vibrational property predicted by the DP model, we report in Fig.~\ref{phonon} the phonon spectra of $P2_1/c$, $Pbca$ and $Pca2_1$ phases. 
An accurate prediction of the phonon spectra requires a good description of the second-order derivative information around local minima of different phases, which is not explicitly considered in the DP-GEN process.
We observe a fairly good agreement between DP and DFT results. 
Adding perturbed structures for calculating the phonon spectra to the training dataset should further improve the DP prediction of this property. 

\subsection{Phase Transitions}
\begin{figure}[ht]
\centering
\includegraphics[scale=1.5]{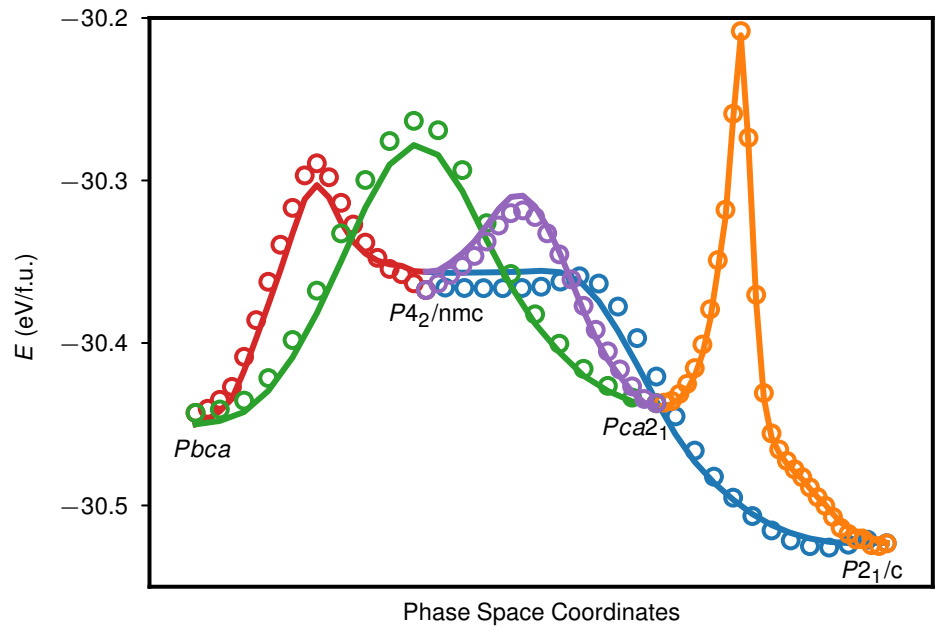}
 \caption{ Comparison of phase transition barriers predicted by DFT (solid line) and DP (empty circle).}
  \label{NEB}
 \end{figure}

The formation of ferroelectric $po$-phase in HfO$_2$ thin films was suggested to have a strong kinetic contribution that the transformation from the metastable $t$- and $po$-phases to the most stable $m$-phase are suppressed by a kinetic barrier~\cite{Park17p9973, Park18p716, Liu19p054404}. In order to use MD to study phase transitions at finite temperatures, it is necessary for the force field to accurately predict the solid-solid phase transition barriers. This is a challenging task as the intermediate structures during the transition are often strongly distorted relative to equilibrium structures. Following a similar protocol established in a previous study~\cite{Liu19p054404}, we first used variable-cell nudged elastic band (VC-NEB) technique to determine the minimum energy paths (MEPs) connecting different phases of HfO$_2$ using the \texttt{USPEX} code~\cite{Oganov06p244704,Lyakhov13p1172,Oganov11p227}. The {\em ab initio} calculations of force and stress tensors were performed using PBE exchange-correlation functional, consistent with the method used to label structures in the DP-GEN scheme. Specifically, five solid-solid phase transitions relevant to the growth of ferroelectric HfO$_2$ thin films were studied: $P2_1/c \leftrightarrow P4_2/nmc $, $P2_1/c \leftrightarrow Pca2_1 $, $Pca2_1 \leftrightarrow P4_2/nmc $,  $Pbca \leftrightarrow P4_2/nmc $, and $Pca2_1 \leftrightarrow Pbca $. The energies of structures of identified MEPs were then evaluated with the DP model. Figure~\ref{NEB} compares the DP and DFT energies along the MEPs, showing excellent agreement between DP and DFT with a MAE of 2.2 meV/atom.

\begin{figure}[ht]
\centering
\includegraphics[scale=1.0]{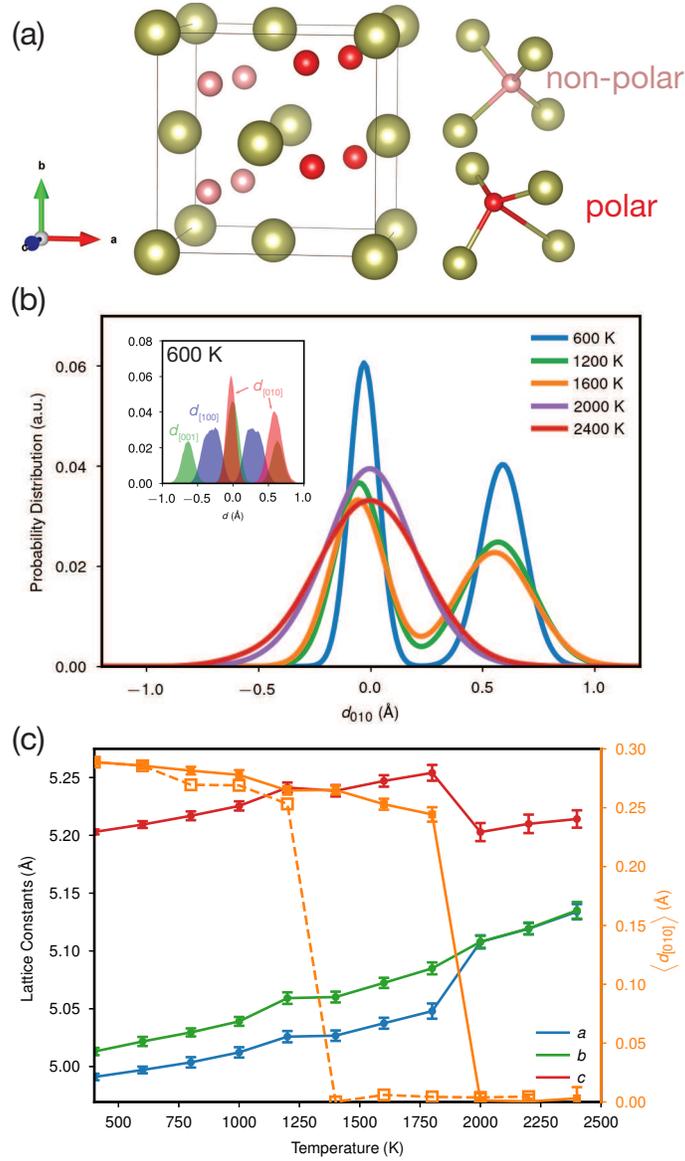}
 \caption{(a) Structure of ferroelectric HfO$_2$ in the space group $Pca2_1$.  Hafnium atoms are denoted by golden balls. Oxygen atoms with zero and non-zero [010] displacements relative to the center of their surrounding Hf$_4$ tetrahedron are colored in pink and red, respectively. 
 (b) Probability distributions of atomic displacements ($d_{[010]}$)  of O atoms along [010] at various temperatures. The inset shows the distributions of O atomic displacements along [100], [010], and [001], respectively, at 600 K. Oxygen atoms have net displacements along [010]. (c) Temperature-dependent lattice constants and average value of $d_{[010]}$ of oxygen atoms from DPMD simulations with a 6144-atom supercell. The dashed organe line is obtained using a 96-atom supercell.}
  \label{feTransition}
 \end{figure}

One major focus of this work is to enable MD simulations of the newly discovered ferroelectric HfO$_2$. We further simulate the temperature-driven phase transitions starting with the ferroelectric $Pca2_1$ phase using the DP model and a 6144-atom supercell. 
The local displacement of the oxygen atom relative to the center of its surrounding Hf$_4$ tetrahedron is used to gauge the local symmetry breaking (Fig.~\ref{feTransition}a) . Figure~\ref{feTransition}b shows the temperature dependence of probability distributions of local oxygen displacements along Cartesian axes. We find that at 600~K, the distributions along the [100] and [001] directions are symmetric, whereas the distribution along [010] is asymmetric with one peak centered around zero and another peak centered around 0.6~\AA~(Fig.~\ref{feTransition}b inset). This is consistent with the structural origin of ferroelectricity in $Pca2_1$ HfO2 that only half of oxygen atoms are locally displaced along the [010] direction (Fig.~\ref{feTransition}a). With increasing temperature, the positive peak of $d_{[010]}$ distribution shifts toward a lower value, indicating a decrease of total polarization and a displacive phase transition. In the high temperature paraelectric phase (2400 K), the $d_{[010]}$  distribution becomes a single peak. Figure~\ref{feTransition}c shows the temperature dependence of lattice constants and the average value of $d_{[010]}$, which clearly reveals a ferroelectric-to-paraelectric phase transition with the tetragonal $P4_2/nmc$ phase being the non-polar high-temperature phase, agreeing with experimental observations. 

It is well known that the phase transition temperature ($T_c$) predicted with MD will suffer from the supercell size effect. The ferroelectric-paraelectric $T_c$ for single-crystal ferroelectric HfO$_2$ obtained using a 96-atom supercell is $\approx$1200~K, comparable with previous {\em ab inito} MD simulations using a supercell of the same size~\cite{Liu19p034032,Fan19p21743}. We confirm that simulations with 6144-atom and 12000-atom supercells yield similar $T_c$ of $\approx$2000~K. This highlights the importance of using a large supercell to obtain the intrinsic $T_c$ for ferroelectric HfO$_2$. 

\subsection{Developing force field beyond pure HfO$_2$}
We make a few general comments here regarding the approach to systematically extend the applicability of the DP model of HfO$_2$. It is well known that the ML method works essentially by interpolating the high-dimensional data contained in the training database. Though it is the DP model with optimized network parameters that will be used in practice, we suggest the associated training database is a more fundamental entity which can be used to re-train a new DP model with accuracy similar to the current one or any other ML-based force field if needed. 
Since the training data is generated with expensive first-principle calculations, making the training database available to the public will greatly facilitate the development of new force fields through community efforts. In this work, we focus on developing a DP model for pure HfO$_2$. Given that extrinsic effects such as dopants and defects can strongly affect the ferroelectric properties of HfO$_2$, it is also desirable to have an accurate and efficient model potential that accounts for these extrinsic effects. Thanks to the ability of deep neural work to faithfully represent complex and highly nonlinear PES, it is expected a DP model with improved transferability can be readily developed by (1) adding new structures with dopants/defects of interests and (2) setting up appropriate exploration runs to generate new structures with dopants/defects. In this spirit, we make our final training database available through a public repository \texttt{DP Library}~\cite{DPLibrary}. 

\section{Conclusions}
In summary, we applied Deep Potential Molecular Dynamics method to develop a force field for HfO$_2$ utilizing a concurrent learning scheme called DP-GEN. The force field is a parameterized deep neural network that maps local atomic environment to atomic energy. Using the model deviation of an ensemble of trained DP models as the indicator for fast labeling not only alleviates the burden of human interventions but also significantly reduces the total cost of first-principles calculations needed to obtain an accurate force field. The accuracy and transferability of the force field are confirmed by comparing a wide range of materials properties ({\em e.g.}, elastic constants, EoSs, and phonon spetra) computed with the DP model to {\em ab initio} results. The DP model can also predict accurately the intrinsic solid-solid transition barriers between different polymorphs of hafnia and capture the main features of temperature-driven phase transitions of the newly discovered ferroelectric phase. We expect the developed DP model will be a useful tool to study the kinetic and dynamical properties of ferroelectric HfO$_2$. The development of a high-fidelity force field of HfO$_2$ demonstrated the ability of DPMD and DP-GEN to deal with materials systems consisted of complex transition metal-oxygen bonds. Finally, we suggest that the training database is a more fundamental entity and its easy access by the public will greatly facilitate the development of ML-based force fields.

 \section{Acknowledgments}
JW and SL acknowledge the support from Westlake Foundation. The computational resource is provided by Westlake Supercomputer Center. 
The work of LZ was supported in part by  the Center of Chemistry in Solution and at Interfaces (CSI) funded by the DOE Award de-sc0019394.

\bibliography{SL}

\newpage
\begin{table}[ht]
\centering
\caption{Lattice parameters $(a,b,c)$ at 0~K calculated by DP and DFT. DP vales in bold.
All phases are orthogonal, except for the $P2_1/c$ phase, whose angle parameters are predicted to be $(\alpha,\beta,\gamma)=(90.000^{\circ}, 99.678^{\circ},90.000^{\circ})$ by both DP and DFT.
\label{latt-param}}
\scalebox{0.8}{
\begin{ruledtabular}
\begin{tabular}{c|ccc}
&  $a$ ($\rm{\AA}$)  & $b$ ($\rm{\AA}$)  & $c$ ($\rm{\AA}$)   \\
\hline
\multirow{2}{*} {\hspace{2em}$P2_1/c$\hspace{2em}} &
5.138 &5.190 &5.322\\
& {\bf 5.146} &{\bf 5.154} &{\bf 5.352} \\
\hline
\multirow{2}{*} {\hspace{2em}$Pbca$\hspace{2em}}  & 5.266 & 10.093  & 5.077 \\
&{\bf 5.265} & {\bf 10.094} & {\bf 5.078} \\
\hline
\multirow{2}{*} {\hspace{2em}$Pca2_1$ \hspace{2em}}  & 
5.266& 5.047 &  5.077\\
 & {\bf 5.265} & {\bf 5.047}& {\bf  5.078}\\
  \hline
\multirow{2}{*} {\hspace{2em}$P2_12_12$ \hspace{2em}}  & 
5.162&  5.181&  4.920\\
 & {\bf 5.153} & {\bf 5.230}& {\bf 4.956}\\
  \hline
\multirow{2}{*} {\hspace{2em}$Pbcn$ \hspace{2em}}  & 
4.850 & 5.833 & 16.032 \\
 & {\bf 4.824} & {\bf 5.839}& {\bf  16.032}\\
  \hline
\multirow{2}{*} {\hspace{2em}$Pmn2_1$ \hspace{2em}}  & 
3.434 & 5.179 & 3.795 \\
 & {\bf 3.456} & {\bf 5.254}& {\bf  3.632}\\
 \hline
  \multirow{2}{*} {\hspace{2em}$P4_2/nmc$ \hspace{2em}}  & 
5.074 & 5.074 &  5.228 \\
 & {\bf 5.075} & {\bf 5.075} & {\bf 5.279} \\
 \hline
\multirow{2}{*} {\hspace{2em}$Fm\bar{3}m$ \hspace{2em}}  & 
5.071 & 5.071 & 5.071\\
 & {\bf 5.067} & {\bf 5.067}& {\bf  5.067}\\
\end{tabular}
\end{ruledtabular}
} 
\end{table}

\newpage
\begin{table}[ht]
\centering
\caption{Elastic constants ($C$), bulk modulus ($B_v$), shear modulus ($G_v$), and Young's modulus ($E_v$) in GPa calculated by DP and DFT. DP vales in bold.\label{Telas}}
\scalebox{0.8}{
\begin{ruledtabular}
\begin{tabular}{c|cccccccc}
& $P2_1/c$ & $Pbca$ & $Pca2_1$ & $P2_12_12$ & $Pbcn$  & $Pmn2_1$ & $P4_2/nmc$ & $Fm\bar{3}m$ \\
\hline
\multirow{2}{*} {\hspace{2em}{$C_{11}$ }\hspace{2em}} &337.59 & 341.01 & 341.78 & 212.55 & 255.83 & 371.37 & 366.52 & 566.83\\
 & {\bf 371.63 }  & {\bf 340.64 }  & {\bf 340.63 }  & {\bf 273.39 }  & {\bf 214.19 }  & {\bf 340.66 }  & {\bf 366.00 }  & {\bf 571.17} \\
\hline
\multirow{2}{*} {$C_{22}$} &390.90 & 395.94 & 395.93 & 212.52 & 298.81 & 351.82 & 366.51 & 566.77\\
 & {\bf 378.18 }  & {\bf 398.00 }  & {\bf 398.00 }  & {\bf 307.07 }  & {\bf 268.04 }  & {\bf 281.57 }  & {\bf 366.00 }  & {\bf 571.17} \\
\hline
\multirow{2}{*} {$C_{33}$} &289.95 & 390.88 & 390.98 & 335.16 & 358.22 & 338.51 & 283.58 & 566.86\\
 & {\bf 369.25 }  & {\bf 373.00 }  & {\bf 372.99 }  & {\bf 357.91 }  & {\bf 344.82 }  & {\bf 335.12 }  & {\bf 241.33 }  & {\bf 571.17}\\
\hline
\multirow{2}{*} {$C_{12}$} &165.01 & 129.59 & 130.43 & 225.53 & 167.25 & 98.31 & 233.50 & 96.60 \\
 & {\bf 160.44 }  & {\bf 129.32 }  & {\bf 129.32 }  & {\bf 141.66 }  & {\bf 148.92 }  & {\bf 62.80 }  & {\bf 227.72 }  & {\bf 99.52} \\
\hline
\multirow{2}{*} {$C_{13}$} &104.46 & 95.06 & 95.38 & 166.47 & 152.60 & 252.60 & 60.08 & 96.60\\
 & {\bf 131.64 }  & {\bf 81.23 }  & {\bf 81.23 }  & {\bf 157.19 }  & {\bf 141.12 }  & {\bf 285.71 }  & {\bf 59.18 }  & {\bf 99.52}\\
\hline
\multirow{2}{*} {$C_{23}$} &162.15 & 126.44 & 126.60 & 142.04 & 126.20 & 176.57 & 60.07 & 96.54\\
 & {\bf 160.53 }  & {\bf 124.77 }  & {\bf 124.77 }  & {\bf 166.95 }  & {\bf 108.44 }  & {\bf 112.87 }  & {\bf 59.18 }  & {\bf 99.52}\\
\hline
\multirow{2}{*} {$C_{44}$} &81.51 & 86.31 & 86.40 & -71.27 & 92.58 & -0.39 & 7.87 & 72.26\\
 & {\bf 104.31 }  & {\bf 88.65 }  & {\bf 88.65 }  & {\bf -15.38 }  & {\bf 85.26 }  & {\bf 19.90 }  & {\bf 33.58 }  & {\bf 65.08} \\
\hline
\multirow{2}{*} {$C_{55}$} &94.55 & -31.24 & -31.17 & -69.80 & 119.52 & 161.71 & 7.87 & 72.25\\
 & {\bf 89.49 }  & {\bf -29.56 }  & {\bf -29.56 }  & {\bf -64.77 }  & {\bf 110.51 }  & {\bf 182.11 }  & {\bf 33.58 }  & {\bf 65.08}\\
\hline
\multirow{2}{*} {$C_{66}$} &126.44 & 109.07 & 109.54 & 127.28 & 129.09 & 74.37 & 169.99 & 72.26\\
 & {\bf 116.85 }  & {\bf 115.00 }  & {\bf 115.00 }  & {\bf 125.49 }  & {\bf 131.31 }  & {\bf 48.54 }  & {\bf 182.65 }  & {\bf 65.08}\\
\hline
\multirow{2}{*} {$B_v$} &221.30 & 203.78 & 204.14 & 205.83 & 200.65 & 235.29 & 191.30 & 253.33\\
 & {\bf 224.69 }  & {\bf 198.47 }  & {\bf 198.47 }  & {\bf 210.68 }  & {\bf 180.48 }  & {\bf 208.35 }  & {\bf 185.75 }  & {\bf 254.77 }\\
\hline
\multirow{2}{*} {$G_v$} &137.40 & 111.58 & 81.34 & 12.32 & 84.61 & 99.36 & 84.71 & 82.75\\
 & {\bf 124.62 }  & {\bf 106.26 }  & {\bf 91.78 }  & {\bf 40.57 }  & {\bf 86.57 }  & {\bf 93.99 }  & {\bf 86.57 }  & {\bf 83.18 }\\
\hline
\multirow{2}{*} {$E_v$} &349.09 & 286.57 & 213.74 & 36.24 & 222.97 & 255.85 & 223.24 & 222.20\\
 & {\bf 321.44 }  & {\bf 275.36 }  & {\bf 236.40 }  & {\bf 114.38 }  & {\bf 226.75 }  & {\bf 240.26 }  & {\bf 226.75 }  & {\bf 220.22 }\\
\end{tabular}
\end{ruledtabular}
} 
\end{table}

\newpage
\clearpage
\newpage
\newpage

\newpage

\newpage

\newpage
\newpage

\newpage
\newpage

\newpage
\newpage

\newpage
\newpage

\end{document}